# Significant improvement of lower critical field in Y doped Nb: potential replacement of basic material for the radio-frequency superconducting cavity

Wei Xie, Yu-Hao Liu, Xinwei Fan, Hai-Hu Wen*

National Laboratory of Solid State Microstructures and Department of Physics, Collaborative Innovation Center of Advanced Microstructures, Nanjing University, Nanjing 210093, China.
**E-mail**:
hhwen@nju.edu.cn

**Abstract**

The research of high energy and nuclear physics needs high power accelerators, and the superconducting radio-frequency (SRF) cavity is regarded as the engine of them. Up to now, the widely used practical and effective material for making the SRF cavity is pure Nb. The key parameter that governs the efficiency and the accelerating field ($E_{acc}$) of a SRF cavity is the lower critical field $H_{c1}$. Here we report a significant improvement of $H_{c1}$ of a new type of alloy, $Nb_{1-x}Y_x$ fabricated by arc melting technique. Experimental investigations with multiple tools including x-ray diffraction, scanning electron microscopy, resistivity and magnetization are carried out, showing that the samples have good quality and a 30-60% enhancement of $H_{c1}$. First principle calculations indicate that this improvement is induced by the delicate tuning of a Lifshitz transition of a Nb derivative band near the Fermi energy, which increases the Ginzburg-Landau parameter and $H_{c1}$. Our results may trigger a replacement of the basic material and thus a potential revolution for manufacturing the SRF cavity.

## 1. Introduction

The SRF cavity was firstly proposed for accelerating charged particles and widely used for the research of high energy and nuclear physics [1]. A lot of research have been conducted on the SRF cavities over the past decades. The performance of a state-of-art SRF cavity is mainly determined by two key parameters [2-4]: the quality factor $Q$ and the accelerating gradient of the electric field $E_{acc}$. The quality factor $Q$ is inversely proportional to the cavity radio-frequency (RF) surface resistance $R_s$ as $Q = G/R_s$ ($G$ is a geometrical factor independent on the cavity material) [2-5]. Generally, it is the energy loss in the skin layer of the cavity that restricts the RF application of superconductors when the RF electromagnetic wave is applied, and this power dissipation is characterized by the surface resistance $R_s$. Thus, $Q$ determines the energy dissipation on the cavity per RF cycle. In principle, $R_s$ is closely related to the normal state resistivity $\rho_n$ (with the Bardeen-Cooper-Schrieffer theory in the dirty limit: $R_s \sim \sqrt{\rho_n}$) and the superconducting gap [2-3]. The former measures the quasiparticle scattering rate and latter influences the excited quasiparticle numbers at a certain temperature. In this case, $Q$ can be evaluated by the normal state resistivity and the superconducting gap [2]. On the other hand, the maximal energy gained during each RF cycle is determined by $E_{acc}$ which is proportional to the peak magnetic field $B_{pk}$ in the way that $E_{acc} = B_{pk}/g$ ($g$ is a factor dependent on the cavity shape, e.g. $g = 4.26$ mT MV$^{-1}$ m for TESLA-shape cavity) [2,4]. In principle, $B_{pk}$ reflects the highest amplitude of the magnetic field for flux to penetrate the interior of the superconducting cavity. Thus, $B_{pk}$ is limited by the maximum field for the superconductor to remain in Meissner state and resist flux penetration. For type-I superconductors ($\kappa < 1/\sqrt{2}$, $\kappa$ is the Ginzburg-Landau (GL) parameter), $B_{pk}$ is limited by the thermodynamic critical field $B_c$, and for type-II superconductors ($\kappa > 1/\sqrt{2}$), $B_{pk}$ is limited by the lower critical field $B_{c1}$ [6,7]. While in practice, due to the existence of surface Bean-Livingston barrier, the superconductor can hold the metastable Meissner state up to the superheating field $B_{sh}$ ($> B_{c1}$) if the surface is very smooth [8-13]. $B_{sh}$ is dependent both on the GL parameter $\kappa$ and the thermodynamic critical field $B_c$. However, $B_{sh}$ is difficult to be reached in practice because of surface roughness of the superconductor [14-16].

Up to now, Nb is still regarded as the most promising material for manufacturing the SRF cavities due to its relatively high critical temperature [2,17], rather high lower-critical field $B_{c1}$ [2,9], large superfluid density among all the element or alloy superconductors, and the high residual resistivity ratio [2,18]. Meanwhile, its high ductility allows for easy manufacturing. However, in practical, the pure Nb cavity is approaching its fundamental limit in terms of the magnetic flux entry field $B_{sh}$ [19-21], and it is highly desired to have materials with better performance than Nb for making the SRF cavity [2,9]. In order to improve the peak magnetic field or quality factor of Nb cavity, some treatments have been carried out, such as the so-called International Linear Collider (ILC) recipe with the combination of the electro-polishing and baking at 120 °C for 48 h [22-25], nitrogen-doping [26-29], nitrogen-infusion [21,30], Nb$_3$Sn coating [31-33], rare earth elements doping [34], etc. Besides, some ideas for achieving higher energy barrier of flux entry were also proposed [2]. But the improvement of $B_{c1}$ due to the aforementioned methods is still very limited. Meanwhile, the multilayer-structure proposed by Gurevich [35,36] and Kubo *et al.* [37-39] has been under extensive research [40,41]. The multilayer-structure consists of a superconducting and an insulating layer coated on the inner surface of the Nb cavity. The thickness of superconducting layer should be in the scale of London penetration depth which is tens of nanometers for Nb [2]. These ideas definitely deserve to be tested, but it is technically very hard for the manufacturing. Other candidate materials for SRF applications have also been explored, such as thin films of Nb [42,43], NbN [3,44], NbTiN [2], FeSe [45], MgB$_2$ [46] and some A15 compounds (Nb$_3$Sn, etc) coated on Nb or Cu substrates [47-49]. Besides the high demand for the quality factor and the peak magnetic field, the candidate materials for SRF applications also need to be tough, easy refreshing and polishing. Thus, we focus on the manipulation of material properties by slightly doping other elements to Nb. Our goal is to increase the lower critical field, but maintain the high $T_c$, large *RRR* and toughness of Nb. After trying doping with many different elements, it is found that the Nb$_{1-x}$Y$_x$ alloys can achieve a lower critical field of about 30-60% higher than Nb. This shows a great potential for the Nb$_{1-x}$Y$_x$ alloys to replace the existing Nb for achieving improved performance of the SRF cavity in the future.

## 2. Experimental details

The Nb$_{1-x}$Y$_x$ alloys were prepared by arc-melting method. Three doping levels were chosen, i.e. $x$=0.05, 0.10, 0.15 for the Nb$_{1-x}$Y$_x$ alloys. The Niobium (99.95%,) and Yttrium (99.9%,) were weighed, ground and pressed into tablets to prepare the precursors according to the corresponding molar ratio in a glove box filled with argon. Then the precursors were melted in the arc-furnace filled with high purity argon. The melting process lasts for at least one minute. To improve the homogeneity of the ingots, all precursors were flipped and re-melted three times. At last, the well-mixed and high quality Nb$_{1-x}$Y$_x$ alloys were obtained. The structural characterization of the Nb$_{1-x}$Y$_x$ and Nb was performed with the x-ray diffraction (XRD) measurements on a Bruker D8 Advanced diffractometer with the Cu-K$\alpha$ radiation. The surface topography and element composition analysis of the Nb$_{1-x}$Y$_x$ and Nb were taken on a Phenom ProX scanning electron microscope (SEM).

For DC magnetization measurements, all the samples were cut into rectangular shape by the wire cutting machine and carefully polished with sandpaper to obtain smooth and flat sample surface. The dimensions and weights of the bulks are 2.93×2.90×0.38 mm$^3$ and 29.63 mg of Nb$_{0.95}$Y$_{0.05}$, 2.93×2.87×0.37 mm$^3$ and 30.60 mg of Nb$_{0.9}$Y$_{0.1}$, 2.93×2.85×0.36 mm$^3$ and 27.97 mg of Nb$_{0.85}$Y$_{0.15}$, 3.00×2.70×0.42 mm$^3$ and 35.25 mg of Nb. The dimensions of Nb$_{1-x}$Y$_x$ and Nb are much close to each other, for the purpose of reducing the effect of demagnetization factor. The DC magnetization measurements were carried out on a SQUID-VSM-7T (Quantum Design). The resistivity measurements were carried out on a physical property measurement system (PPMS 16T, Quantum Design), using the standard four-probe method.

## 3. Results and discussion

Alloys with three nominal doping levels have been prepared, they are Nb$_{0.95}$Y$_{0.05}$, Nb$_{0.9}$Y$_{0.1}$ and Nb$_{0.85}$Y$_{0.15}$. The doping ratio $x$ of the Nb$_{1-x}$Y$_x$ alloys is the mole ratio of the precursor during preparation. A bulk sample of high-purity Nb is used as comparison. Figure 1(a) shows the temperature dependence of magnetization measured in zero-field-cooled (ZFC) mode of Nb$_{1-x}$Y$_x$ and Nb. All samples are cut into a rectangular shape with almost the same sizes to reduce the interference of demagnetization effect [50,51]. The applied magnetic field is 10 Oe and parallel to the lateral plane of the samples, which is defined as ab-plane. The superconducting transition temperature $T_c$ is determined by the point where the magnetization starts to deviate from the paramagnetic background at high temperatures, and the results are summarized in Table 1. The $T_c$ = 9.35 K of $Nb_{0.9}Y_{0.1}$ is a bit higher than $T_c$ = 9.17 K of Nb and the $T_c$ of $Nb_{0.95}Y_{0.05}$ and $Nb_{0.85}Y_{0.15}$ is extremely close. This indicates that the slightly-doping of Y can slightly raise the $T_c$ of Nb, which is consistent with previous results [52], and the operation temperature of SRF applications by using the new alloy is guaranteed. The very steep magnetization transitions in Figure 1(a) indicate a perfect Meissner shielding effect and high quality of all samples. Thus, it is reliable to determine the lower critical field $H_{c1}$ by the deviation from the linear Meissner line of the $M(H)$ curves when the field is parallel to the plane. Figures 1(b)-(d) show the temperature dependence of the electrical resistivity of $Nb_{1-x}Y_x$ with different nominal doping concentrations, respectively. The insets in Figures 1(b)-(d) give details of the resistivity at low temperatures and the inserted images show the samples with electrodes for resistivity measurements. When temperature decreases, the resistivity decreases monotonically and shows a metallic behavior before entering the superconducting state. Meanwhile, when the applied field increases, the transition of the resistivity widens slightly and shifts to the lower temperatures until the superconductivity is suppressed completely. The residual resistivity ratio (*RRR*) is related with the thermal conductivity of the material and important for the SRF cavities. The *RRR* of $Nb_{1-x}Y_x$ is defined as $RRR = \rho(300\ K)/\rho(10\ K)$. The measured results are *RRR* = 14.5 for $Nb_{0.95}Y_{0.05}$, *RRR* = 14.9 for $Nb_{0.9}Y_{0.1}$ and *RRR* = 11.1 for $Nb_{0.85}Y_{0.15}$, respectively. The *RRR* of $Nb_{1-x}Y_x$ are quite large, indicating a good quality of the samples, but these values are smaller than that of high-purity Nb (*RRR* = 243 in Supplementary Figure S4) [18]. This may be attributed to the slightly doping of Y and the increase of impurity scattering. However, with heat treatment at high temperature, the *RRR* of $Nb_{1-x}Y_x$ may be further improved [52]. The temperature dependence of the electrical resistivity of our high-purity Nb is also given in Supplementary Figure S4.

The crystal structures of $Nb_{1-x}Y_x$ and Nb are examined by x-ray diffraction (XRD) and the XRD patterns are shown in Figure 2(a). The Nb has a cubic (bcc) symmetry and a space group of Im -3 m (number 229) [53]. The $Nb_{1-x}Y_x$ have the same index peaks with Nb and no other peaks can be observed in Figure 2(a). This indicates that the $Nb_{1-x}Y_x$ have the same crystal structure with Nb and the slightly doping of Y does not change the structure of Nb significantly. However, the (110) peaks of XRD pattern of $Nb_{1-x}Y_x$ show a slightly shift to a higher angle compared with Nb in the inset of Figure 2(a). Meanwhile, the lattice parameters *a* calculated from the XRD patterns are *a* = 3.3065(3) Å for $Nb_{0.95}Y_{0.05}$, *a* = 3.3027(8) Å for $Nb_{0.9}Y_{0.1}$ and *a* = 3.3095(9) Å for $Nb_{0.85}Y_{0.15}$, compared with *a* = 3.3167(5) Å of Nb. The slightly decrease of lattice parameters and the shift of (110) peaks from XRD patterns both indicate that a measurable amount of Y has been successfully incorporated into Nb. The energy dispersive spectrums (EDS) of $Nb_{1-x}Y_x$ and Nb from scanning electron microscope (SEM) measurements are shown in Figure 2(b). The corresponding SEM images are given in Supplementary Figure S5. The SEM image in Figure 2(b) shows a quite smooth surface of $Nb_{0.9}Y_{0.1}$ with no clear grain boundary. The inset in Figure 2(b) shows the enlarged view of the EDS around 1.9 keV and the dashed vertical lines are the characteristic peaks of Nb (olive) and Y (pink) in the vicinity. With increasing ratio of Y in the alloys, the peak around 1.9 keV is elevated and shifts to higher energy, which is indicated by the black arrow in the inset of Figure 2(b). This confirms the existence of Y in the alloys and the atomic compositions of Y from EDS analyses are 1.14% for $Nb_{0.95}Y_{0.05}$, 1.82% for $Nb_{0.9}Y_{0.1}$ and 5.46% for $Nb_{0.85}Y_{0.15}$. This shows a loss of Y in the $Nb_{1-x}Y_x$ alloys compared with the nominal composition of starting materials and may be attributed to a saturation of solubility of Y in Nb, which is consistent with previous results [52]. Even with this solution limit, the increase of the lower critical field of the alloys can be clearly seen.

Figures 3(a)-(c) show the isothermal magnetic-hysteresis-loops (MHLs) of $Nb_{1-x}Y_x$ and Nb measured at the same temperature, respectively. The temperature ranges from 2 K to 6 K and the magnetic field ranges from -1 T to 1 T. The field is parallel to the ab-plane of the samples, for the purpose of reducing the influence of demagnetization effect [50,51]. The MHLs of $Nb_{1-x}Y_x$ show stronger symmetric feature with respect to the horizontal line than that of Nb, and this indicates a stronger bulk flux pinning of the alloy [51]. The avalanche-like flux jump effect with the flux abruptly entering the sample can be seen in $Nb_{1-x}Y_x$, which is closely related to the thermomagnetic instabilities of the critical state, and it becomes more obvious at low temperatures [54,55]. However, this "harmful" flux jump only occurs when the vortices begin to penetrate the sample, thus it may not affect the SRF applications which work mainly in the Meissner state [2,4,9]. On the other hand, the flux trapped during the cooling process significantly degrades the quality factor *Q*. Understanding the tendency of flux trapping in $Nb_{1-x}Y_x$ and its sensitivity is one of the tasks for the future [56,57]. The insets in Figures 3(a)-(c) present details of the Meissner state of the initial $M(H)$ curves in the low-field region. It should be pointed out that, the magnetization has been corrected to make $M(H)$ curves of the Meissner state of different samples merge into one line. Through comparison of the initial $M(H)$ curves, it can be seen that the $Nb_{1-x}Y_x$ can hold the Meissner state (the linear part of $M(H)$ curves) to a higher field than Nb, and among them the $Nb_{0.9}Y_{0.1}$ sample shows the best performance. The lower critical field $H_{c1}$ is defined as where the $M(H)$ curve starts to deviate from the linear line. To obtain the value of $H_{c1}$ precisely, the deviation from the linear Meissner line of the $M(H)$ curve has been calculated accurately for each sample [58,59]. Firstly, the linear Meissner line is obtained by fitting to the linear part of $M(H)$ curve in the low-field region. Because of the small difference in sample size, the linear Meissner lines of $Nb_{1-x}Y_x$ and Nb are slightly different. Secondly, we subtract the corresponding Meissner lines from the initial $M(H)$ curves and the deviation of magnetization can then be obtained. The same criterion of $\Delta M = 5$ emu cm$^{-3}$ is used to determine the value of $H_{c1}$ for all temperatures, it is comparable to the background signal. The deviation of magnetization at 2 K of $Nb_{1-x}Y_x$ and Nb are shown in Figure 3(d) and the red horizontal line is the criterion of magnetization for determining the $H_{c1}$. The obtained results from Figure 3(d) are $H_{c1}$(2 K) = 1515 Oe of $Nb_{0.95}Y_{0.05}$, $H_{c1}$(2 K) = 1953 Oe of $Nb_{0.9}Y_{0.1}$, $H_{c1}$(2 K) = 1509 Oe of $Nb_{0.85}Y_{0.15}$ and $H_{c1}$(2 K) = 1247 Oe of Nb. The $H_{c1}$ of $Nb_{0.9}Y_{0.1}$ is 57% higher than that of Nb and gives rises to the strongest enhancement, and all three alloys show a clear improvement of $H_{c1}$ compared with Nb. The determination of $H_{c1}$ at other temperatures are given in Supplementary Figure S1 and Supplementary Table S1.

Figure 4(a) shows the temperature dependence of the lower critical field $H_{c1}$ of $Nb_{0.9}Y_{0.1}$, $Nb_{0.85}Y_{0.15}$ and Nb. In order to obtain the values of $H_{c1}$ at zero temperature, the empirical formula $H(T) = H(0)[1-(T/T_c)^2]^n$ is used to fit the $H_{c1}$(T) of all samples and the fitting curves are shown in Figure 4(a), respectively. The fitting parameters are $n = 1.00$ of $Nb_{0.9}Y_{0.1}$, $n = 0.98$ of $Nb_{0.85}Y_{0.15}$ and $n = 1.20$ of Nb, and summarized in Table 1. $H_{c1}$(0) = 2055 Oe of $Nb_{0.9}Y_{0.1}$ is the highest value and 62% higher than $H_{c1}$(0) = 1267 Oe of Nb (Table 1). Concerning the values of $H_{c1}$(0) for pure Nb reported in literatures [2,9,60], there are clear discrepancies. The often cited value of $H_{c1}$(0) =1700 Oe was adopted from Reference. 60, while we find that the authors there used the fully penetration field on the MHL as the $H_{c1}$, not the deviating point of the $M(H)$ curve from the linear Meissner line. In another report with the samples of Nb single crystals, the $H_{c1}$ at 1.83K is smaller than 1235 Oe. This is close with our values. If taking the fully penetration field as $H_{c1}$, that would give a value beyond 2500 Oe ($T = 2$ K) for the sample $Nb_{0.9}Y_{0.1}$. We believe that an appropriate post-annealing of our Nb samples may increase the $H_{c1}$(0) value further. But in any case, this value is smaller than that in the Y doped samples. For the Nb samples with polished surfaces, we find that the $H_{c2}$(0) values determined from the resistive onset transition temperature is quite high ($\geq 1.5$ T) and much larger than that from other literatures [2,8,42], we believe that is probably due to the surface superconductivity. Thus we use the temperature dependent magnetization which usually reflects a bulk property to determine the upper critical field $H_{c2}$ (Supplementary Figure S2-S3). The empirical formula $H(T) = H(0)[1-(T/T_c)^2]^n$ is used to fit the $H_{c2}$ of all samples to extrapolate $H_{c2}$(0) (Supplementary Figure S3). The $H_{c2}$(0) of $Nb_{1-x}Y_x$ are twice that of Nb (Table 1). The clear enhancement of $H_{c1}$ in Y-doped Nb is of great significance for the application of the alloy. Above all, the new alloys $Nb_{1-x}Y_x$ can serve as a promising candidate material for SRF applications.

In order to study the physical reason of the improved $H_{c1}$ in $Nb_{1-x}Y_x$ compared with Nb, we carry out the first principle calculation. Figure 5(a) illustrates the calculated band structure of Nb. There are both the hole and electron pockets near the Γ point of the band structure of Nb. When the doping level of the alloy is changed, the Fermi level moves up and down and a Lifshitz transition can occur. This Lifshitz transition occurs most likely for the band between N-Γ in which a band bottom appears very close to the Fermi energy. Figure 5(b) illustrates the total density of states (DOS) of Nb, which consists of the DOS of both $p$ orbit and $d$ orbit. It can be seen that, the DOS of $d$ orbit accounts for the main part of the total DOS of Nb. There is a sharp peak of DOS at slightly negative energy side. This is induced by the shallow band bottom between N-Γ. If doping with holes in the Nb (the substitution of Nb by Y is hole doping), the Fermi energy will drop down (the zero energy point moves left in Figure 5(b)), and thus the DOS will increase to a higher value (the sharp peak on the left of zero energy point in Figure 5(b)). Figure 5(c) illustrates the three-dimensional (3D) Fermi surface of Nb with color-coded Fermi velocities, and the first Brillouin zone (BZ) of Nb has a symmetry of dodecahedron. In principle, the coherence length $\xi$ can be written as $\xi = \hbar v_F/\pi\Delta$, $v_F$ is the Fermi velocity, $\Delta$ is the superconducting gap and $\hbar$ is the reduced Planck constant. The superconducting gap remains almost the same for the alloys compared with Nb and so is $\xi$. On the other hand, the condensed carrier density $n_s/m^*$ can be written as $n_s/m^* = \sigma/e^2\tau \propto \sigma/\tau$, $n_s$ is the superconducting carrier density, $m^*$ is the effective mass, $\sigma$ is the conductivity and $\tau$ is the relaxation time. As illustrated in Figure 5(d), the condensed carrier density of Nb decreases with slightly doping of Y into Nb (hole-doping). The London penetration depth $\lambda$ can be written as $\lambda = (m^*/\mu_0 n_s e^2)^{1/2}$ and therefore increases slightly [61]. For these reasons, the GL parameter $\kappa = \lambda/\xi$ increases.

In the case of small $\kappa$ value, the vortex core should be taken into consideration in evaluating the vortex tension energy and the lower critical field. Thus, through theoretical calculation given by Brandt [62,63], the $\kappa$ dependence of the ratio of critical field is $H_{c1}/H_{c2} =[\ln\kappa + \alpha(\kappa)]/(2\kappa^2)$, $\alpha(\kappa) = 0.5+\exp[-0.415-0.775 \ln\kappa -0.13(\ln\kappa)^2]$. The curve is shown in Figure 4(b) and the $\kappa$ ranges from 0.75 to 10. The above formulae is valid when $T$ is close to $T_c$. Therefore, we use the $H_{c1}$ and $H_{c2}$ at 6 K to calculate the value of $\kappa$. The results are $\kappa = 3.09$ of $Nb_{0.95}Y_{0.05}$, $\kappa = 2.49$ of $Nb_{0.9}Y_{0.1}$, $\kappa = 2.88$ of $Nb_{0.85}Y_{0.15}$ and $\kappa = 2.25$ of Nb. The thermodynamic critical field $H_c$ is given [62,63] by $H_{c1}/H_c = [\ln\kappa + \alpha(\kappa)]/(\sqrt{2}\kappa)$ in Figure 4(b) and the superheating field $H_{sh}$ is determined by $H_{sh}$ =1.26 $H_c$ ($\kappa \approx 1$) [12]. For the samples studied here, we have calculated different quantities and the results are summarized in Table 1. The $\kappa$ of Nb$_{1-x}$Y$_x$ are higher than that of Nb (Table 1), and the latter is higher than the values for Nb from other literatures ($\kappa$ ~1) [2,9]. To deal with this difference, we use the fully penetration field as the criteria of $H_{c1}$ as adopted by others [60], and we have $H_{c1}$(6 K) = 955 Oe of Nb. With the upper critical field $H_{c2}$(6 K) = 0.43 T of Nb and the Brandt's formula for $H_{c1}/H_{c2}$, we have $\kappa$ = 1.80 of Nb. This value of $\kappa$ is much close to 1. The *RRR* value is often used to estimate $\kappa$ of Nb and we have $\kappa/\kappa_0$ = 1.06 for *RRR* = 243 in our experiment, where $\kappa_0$ is the GL parameter for the pure material [64]. The value 1.06 is close to that of ideal pure Nb [64]. The $H_c$(0) of Nb is close to the values from other literatures ($H_c$(0) ~ 0.2 T) [65,66], which reflects the bulk properties of superconductors. With the slight Y doping, the $H_{c2}$ of Nb$_{1-x}$Y$_x$ is elevated a lot compared with the pure Nb (Table 1). Due to the both increase of $H_{c2}$ and $\kappa$, the lower critical field $H_{c1}$ is increased as well. Among all the alloys, Nb$_{0.9}$Y$_{0.1}$ gives rises to the highest value of $H_{c2}$(0) (=1.71 T) and an increase of $\kappa$ (=2.49), and therefore the highest improvement of $H_{c1}$(0) (=2055 Oe). This is consistent with our theoretical argument. Again, here we want to emphasize that, the $H_{c1}(T)$ here are all determined from the very first deviating point of the $M(H)$ curve from the Meissner linear line. Meanwhile, $H_{sh}$(0) of Nb$_{1-x}$Y$_x$ are higher than that of Nb in our study, and approach that of Nb$_3$Sn [2,9,38]. All these indicate a promising prospect of Nb$_{1-x}$Y$_x$ for SRF applications.

## 4. Conclusions

We report systematic investigations on a new type of alloy Nb$_{1-x}$Y$_x$ prepared by arc melting method. The lower critical field $H_{c1}$ of Nb$_{1-x}$Y$_x$ is found to be 30-60% higher than that of high-purity Nb. This may greatly improve the accelerating gradient of the electric field $E_{acc}$ for a superconducting cavity. Thus the Nb$_{1-x}$Y$_x$ alloy may serve as a promising candidate for replacing Nb in the SRF cavity applications.

## Data Availability Statement

The data that support the findings of this study are available upon reasonable request from the authors.

## Acknowledgements

We appreciate the useful discussions with Peng Sha and Chao Dong in Institute of High Energy Physics, CAS. We would also like to thank Alex Gurevich at Old Dominion University for useful discussions. This work is supported by the National Natural Science Foundation of China (Grants No. A0402/11927809, No. A0402/11534005), National Key R and D Program of China (Grant No. 2022YFA1403201), and the Strategic Priority Research Program of Chinese Academy of Sciences (Grant No. XDB25000000).

## References

[1] Banford A P and G H Stafford 1961 *J. Nucl. Energy C.* **3** 287
[2] Valente- Feliciano A 2016 *Supercond. Sci. Technol.* **29** 113002
[3] Antoine C Z *et al* 2019 *Supercond. Sci. Technol.* **32** 085005
[4] Gurevich A 2017 *Supercond. Sci. Technol.* **30** 034004
[5] Gurevich A and Kubo T 2017 *Phys. Rev. B.* **96** 184515
[6] Abrikosov A A 1957 *J. Exp. Theor. Phys. U. S. S. R.* **32** 1442
[7] Ginzburg V L and Landau L D 1950 *Zh. Eksp. Teor. Fiz.* **20** 1064
[8] Bean C P and Livingston J D 1964 *Phys. Rev. Lett.* **12** 14
[9] Liarte D B *et al* 2017 *Supercond. Sci. Technol.* **30** 033002
[10] Lin F P-J and Gurevich A 2012 *Phys. Rev. B.* **85** 054513
[11] Kubo T 2020 *Phys. Rev. Research* **2** 033203
[12] Dolgert A J *et al* 1996 *Phys. Rev. B.* **53** 5650
[13] Transtrum M K *et al* 2011 *Phys. Rev. B* **83** 094505
[14] Buzdin A *et al* 1998 *Physica C* **294** 257
[15] Aladyshkin A Yu *et al* 2001 *Physica C* **361** 67
[16] Kubo T 2015 *Progress of Theoretical and Experimental Physics* **2015** 063G01
[17] Bieler T R *et al* 2010 *Phys. Rev. ST Accel. Beams* 13 031002
[18] Liu B *et al* 2021 *Nuclear Inst. and Methods in Physics Research* **A993** 165080
[19] Geng R L *et al* 2007 *Proceedings of PAC07, Albuquerque, New Mexico, USA (JACoW, CERN Geneva, 2007)* p. 2337
[20] Kubo T *et al* 2014 *Proceedings of IPAC2014, Dresden, Germany (JACoW, CERN Geneva, 2014)* p. 2519
[21] Grassellino A *et al* 2017 *Supercond. Sci. Technol.* **30** 094004
[22] Saito K *et al* 1997 *Proceedings of SRF1997, Abano Terme, Padova, Italy (JACoW, CERN Geneva, 1997)* p. 795
[23] Lilje L *et al* 1999 *Proceedings of SRF1999, La Fonda Hotel, Santa Fe, New Mexico, USA (JACoW, CERN Geneva, 1999)* p. 74
[24] Romanenko A *et al* 2013 *Phys. Rev. ST Accel. Beams* **16** 012001
[25] Romanenko A *et al* 2014 *Appl. Phys. Lett.* **104** 072601
[26] Grassellino A *et al* 2013 *Supercond. Sci. Technol.* **26** 102001
[27] Dhakal P *et al* 2013 *Phys. Rev. ST Accel. Beams* **16** 042001
[28] Yang L *et al* 2021 *Nanotechnology* **32** 245701
[29] Lechner E M *et al* 2020 *Phys. Rev. Applied* **13** 044044
[30] Grassellino A *et al* 2018 *arXiv:*1806.09824
[31] Posen S and Hall D L 2017 *Supercond. Sci. Technol.* **30** 033004
[32] Keckert S *et al* 2019 *Supercond. Sci. Technol.* **32** 075004
[33] Posen S *et al* 2021 *Supercond. Sci. Technol.* **34** 025007
[34] Jiang T *et al* 2014 *Chin. Phys. B* **23** 057403
[35] Gurevich A 2006 *Appl. Phys. Lett.* **88** 012511
[36] Gurevich A 2015 *AIP Advance* **5** 017112
[37] Kubo T, Iwashita Y and Saeki T 2014 *Appl. Phys. Lett.* **104** 032603
[38] Kubo T 2017 *Supercond. Sci. Technol.* **30** 023001
[39] Kubo T 2021 *Supercond. Sci. Technol.* **34** 045006
[40] Ito H *et al* 2019 *Proceedings of SRF2019, Dresden, Germany (JACoW, CERN Geneva, 2019)* p. 632

[41] Katayama R *et al* 2019 *Proceedings of SRF2019, Dresden, Germany (JACoW, CERN Geneva, 2019)* p. 807
[42] Ries R *et al* 2021 *Supercond. Sci. Technol.* **34** 065001
[43] Valente-Feliciano A *et al* 2021 *arXiv:*2204.02536
[44] Leith S *et al* 2021 *Supercond. Sci. Technol.* **34** 025006
[45] Lin Z *et al* 2021 *Supercond. Sci. Technol.* **34** 015001
[46] Tan T *et al* 2016 *Sci. Rep.* **6** 35879
[47] Sharma R G 1987 *Cryogenics* **27** 361–78
[48] Stewart G R 2015 *Physica C* **514** 28–35
[49] Barzi E *et al* 2021 *arXiv:*2203.09718
[50] Prozorov R and Kogan V G 2018 *Phys. Rev. Applied* **10** 014030
[51] Xie W, Liu Y H and Wen H H 2022 *Phys. Rev. B.* **105** 014505
[52] Koch C C and Kroeger D M 1975 *J. Less-Common Met.* **40** 29-38
[53] Seybolt A U 1954 *JOM* **6** 774-776
[54] Evetts J E, Campbell A M and Dew-hughes D 1964 *Philosophical Magazine* **10:104** 339-343
[55] Tsindlekht M I *et al* 2014 *Phys. Rev. B.* **90** 014514
[56] Romanenko A *et al* 2014 *Appl. Phys. Lett.* **105** 234103
[57] Huang S *et al* 2016 *Phys. Rev. Accel. Beams* **19** 082001
[58] Liang R, Bonn D A, Hardy W N and Broun D 2005 *Phys. Rev. Lett.* **94** 117001
[59] Ren C, Wang Z S, Luo H Q, Yang H, Shan L and Wen H H 2008 *Phys. Rev. Lett.* **101** 257006
[60] Finnemore D K, Stromberg T F and Swenson C A 1966 *Phys. Rev.* **149** 231
[61] London F and London H 1935 *Proc. Roy. Soc.* **A155** 71
[62] Brandt E H 2003 *Phys. Rev. B.* **68** 054506
[63] Gurevich A 2012 *Rev. Accel. Sci. Techn.* **05** 119
[64] Ooi S *et al* 2021 *Phys. Rev. B.* **104** 064504
[65] Mcconville T and Serin B 1965 *Phys. Rev.* **140** A1169
[66] Junod A, Jorda J L and Muller J 1986 *J. Low Temp. Phys.* **62** 301

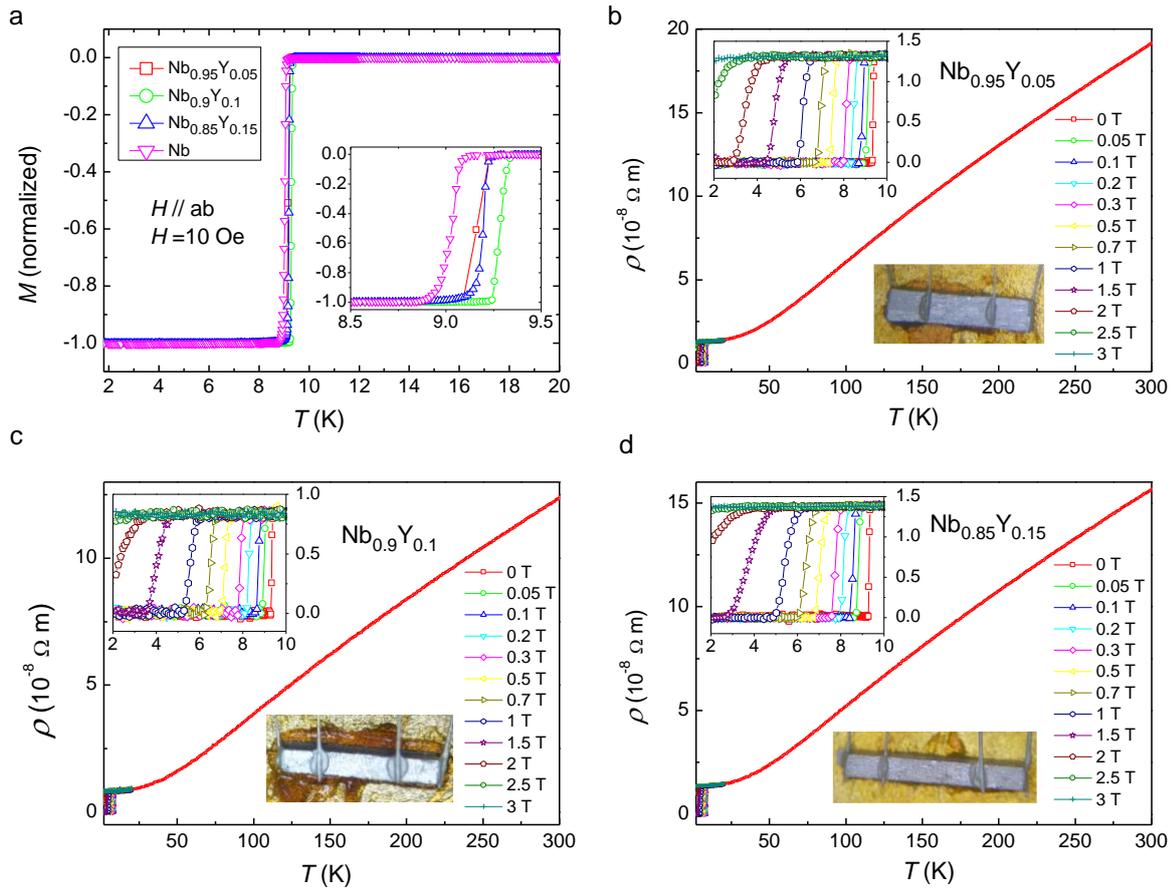

**Figure 1.** Temperature dependence of magnetization and resistive transport properties of the $Nb_{1-x}Y_x$ alloys. (a) Temperature dependence of magnetization measured in zero-field-cooled (ZFC) mode of the $Nb_{1-x}Y_x$ alloys and Nb. The magnetizations of all samples are normalized. Inset shows details of the magnetization curves $M(T)$ in the vicinity of $T_c$. Temperature dependence of resistivity of (b) $Nb_{0.95}Y_{0.05}$, (c) $Nb_{0.9}Y_{0.1}$, (d) $Nb_{0.85}Y_{0.15}$. Insets in (b)-(d) show enlarged views of resistivity at low temperatures. Images in (b)-(d) are the samples with electrodes for resistivity measurement.

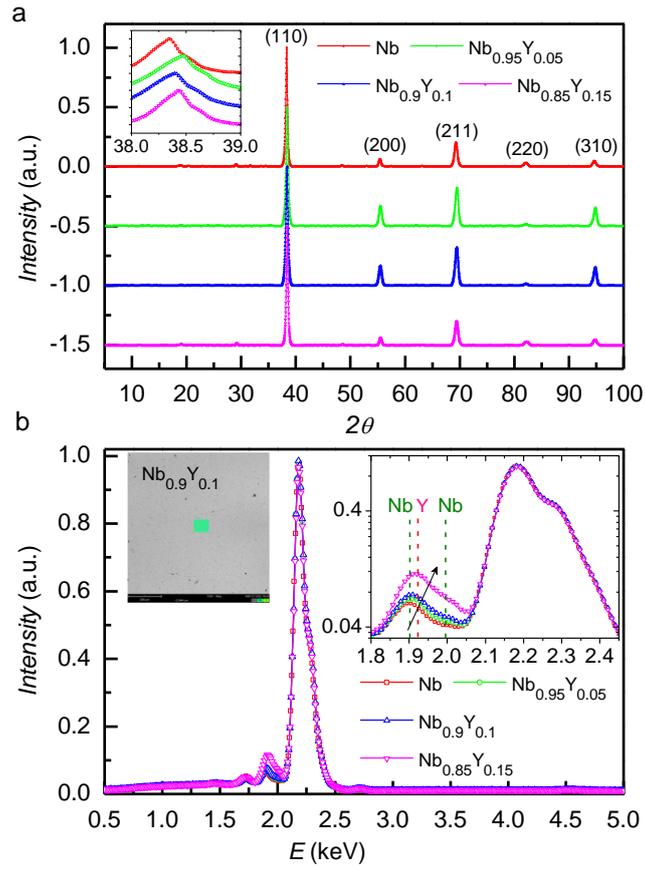

**Figure 2.** XRD, SEM image and energy dispersive spectrums (EDS) of the $Nb_{1-x}Y_x$ alloys and Nb. (a) XRD of the $Nb_{1-x}Y_x$ alloys and Nb. Inset is the enlarged view of (110) peak in XRD patterns. (b) EDS of the $Nb_{1-x}Y_x$ alloys and Nb. Image is SEM image of $Nb_{0.9}Y_{0.1}$ corresponding to the EDS measurement. Inset is the enlarged view of peaks around 1.9 keV in EDS, the dashed vertical lines are the characteristic peaks of Nb (olive) and Y (pink) in the vicinity, and the black arrow shows the elevation and high-energy shift of the peak which is attributed to the existence of Y in the alloys.

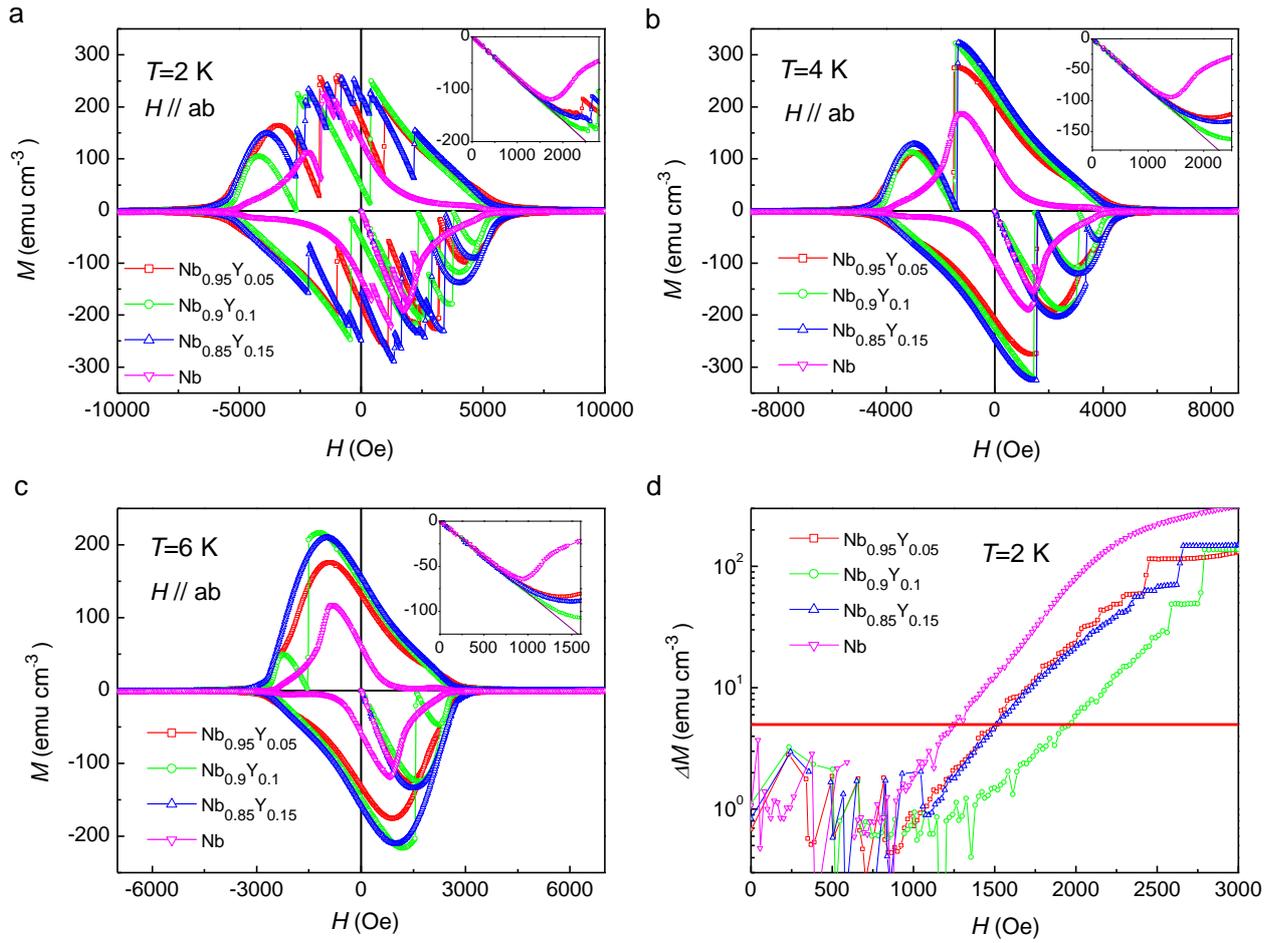

**Figure 3.** MHLs and determination of $H_{c1}$ for the $Nb_{1-x}Y_x$ alloys and Nb. MHLs of the $Nb_{1-x}Y_x$ alloys and Nb at: (a) 2 K, (b) 4 K, (c) 6 K. Insets in (a)-(c) are enlarged views of Meissner state and the initial penetration stage at low fields, the magnetizations of different samples are normalized. (d) The deviation of initial penetration magnetization with respect to the corresponding Meissner line at 2 K, the criterion used to determine $H_{c1}$ is $\Delta M = 5$ emu cm$^{-3}$ (red horizontal line).

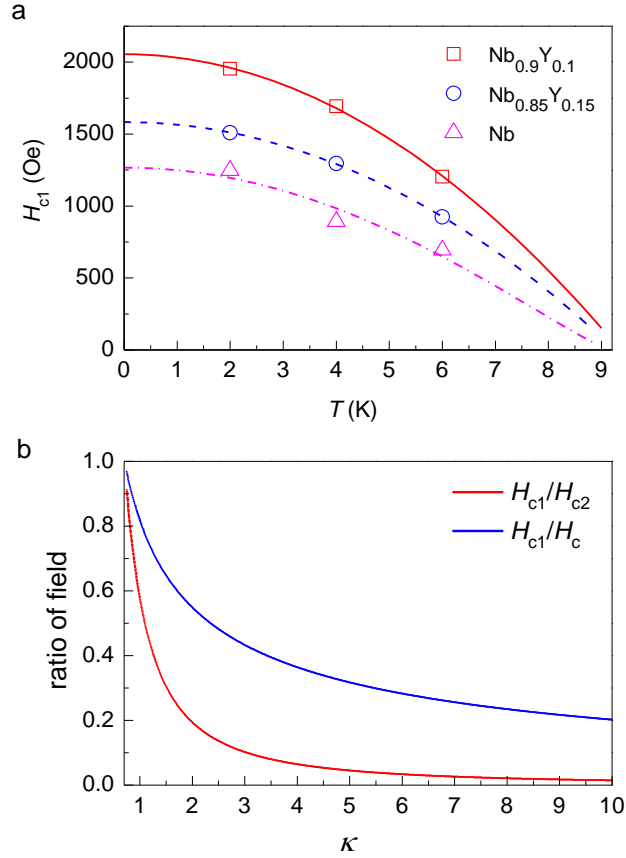

**Figure 4.** The lower critical field $H_{c1}$ of the $Nb_{1-x}Y_x$ alloys and Nb in our experiment. All $H_{c1}$ values are determined from the deviating point of the $M(H)$ curve from the linear Meissner line. (a) Temperature dependence of the lower critical field $H_{c1}$ of the $Nb_{1-x}Y_x$ alloys and Nb, the lines are the corresponding fitting curves using the empirical formula $H(T) = H(0)[1-(T/T_c)^2]^n$. The fitting parameters are $n = 1.00$ of $Nb_{0.9}Y_{0.1}$, $n = 0.98$ of $Nb_{0.85}Y_{0.15}$ and $n = 1.20$ of Nb. (b) The $\kappa$ dependence of the ratio of critical field from theoretical calculation [62,63] given by $H_{c1}/H_{c2} = [\ln\kappa + \alpha(\kappa)]/(2\kappa^2)$ and $H_{c1}/H_c = [\ln\kappa + \alpha(\kappa)]/(\sqrt{2}\kappa)$, $\alpha(\kappa) = 0.5+\exp[-0.415-0.775\ln\kappa-0.13(\ln\kappa)^2]$.

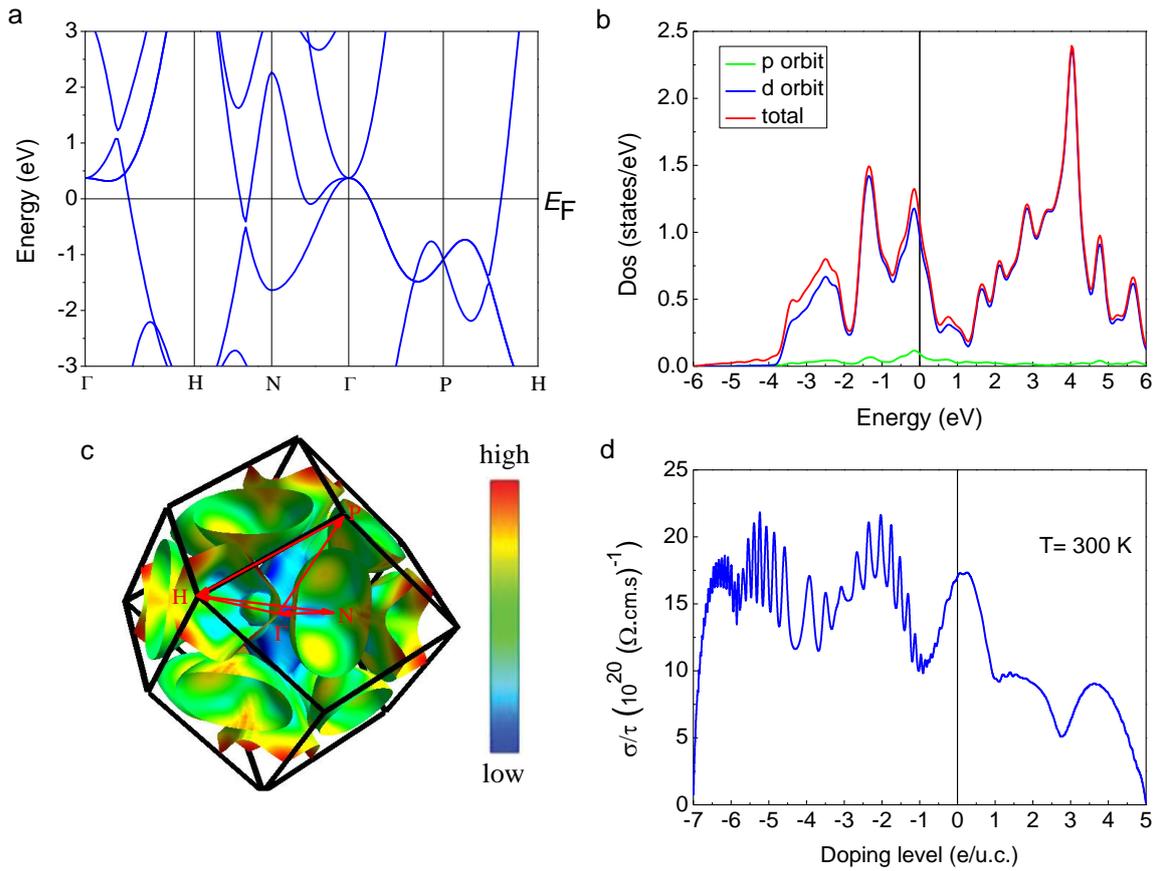

**Figure 5.** First principle calculations of the electronic structure of Nb. (a) Band structure of Nb. (b) Density of states (DOS) of Nb, including the total DOS (red line), the DOS of *p* orbit (green line) and *d* orbit (blue line). (c) 3D Fermi surface of Nb with color-coded Fermi velocities (blue is low and red is high velocity). (d) The ratio of conductivity and relaxation time $\sigma/\tau$ as a function of doping level for Nb at 300 K.

**Table 1.** Superconducting parameters for the $Nb_{1-x}Y_x$ alloys and Nb. Include the critical temperature $T_c$, the lower critical field $H_{c1}(0)$, the upper critical field $H_{c2}(0)$, the Ginzburg-Landau parameter $\kappa$, the thermodynamic field $H_c(0)$ and the superheating field $H_{sh}(0)$. $H_{c1}$ and $H_{c2}$ of different samples are fitted by the empirical formula $H(T) = H(0)[1-(T/T_c)^2]^n$ to obtain $H_{c1}(0)$ and $H_{c2}(0)$ respectively, and the values in the brackets are the corresponding fitting parameters $n$. The Ginzburg-Landau parameter $\kappa$ is calculated [62,63] by $H_{c1}/H_{c2} =[\ln\kappa + \alpha(\kappa)]/(2\kappa^2)$, with $\alpha(\kappa) = 0.5+\exp[-0.415-0.775\ln\kappa-0.13(\ln\kappa)^2]$, and the $H_{c1}$ and $H_{c2}$ at 6 K are uesd in the calculation. The thermal dynamic field is calculated [62,63] by $H_{c1}/H_c = [\ln\kappa + \alpha(\kappa)]/(\sqrt{2}\kappa)$. The superheating field is calculated [2,12] by $H_{sh}=1.26 H_c$ ($\kappa \approx 1$).

| sample parameter | $Nb_{0.95}Y_{0.05}$ | $Nb_{0.9}Y_{0.1}$ | $Nb_{0.85}Y_{0.15}$ | Nb |
|---|---|---|---|---|
| $T_c$ (K) | 9.22 | 9.35 | 9.24 | 9.17 |
| $H_{c1}(0)$ (Oe) | 1557 (1.22) | 2055 (1.00) | 1584 (0.98) | 1267 (1.20) |
| $H_{c2}(0)$ (T) | 1.60 (1.15) | 1.71 (1.27) | 1.52 (1.04) | 0.87 (1.24) |
| $\kappa$ | 3.09 | 2.49 | 2.88 | 2.25 |
| $H_c(0)$ (T) | 0.37 | 0.43 | 0.36 | 0.25 |
| $H_{sh}(0)$ (T) | 0.47 | 0.54 | 0.45 | 0.31 |